\documentclass[twocolumn,showpacs,preprintnumbers,superscriptaddress,amsmath,amssymb,nofootinbib]{revtex4-1}
\input psfig.sty
\usepackage{graphicx}

\begin{document}

\title{Thermodynamic aspects of dark energy fluids}

\author{Ed\'esio M. Barboza Jr.} \email{edesiobarboza@uern.br}
\affiliation{Departamento de F\'isica, Universidade do Estado do Rio Grande do Norte, 59610-210, Mossor\'o, RN, Brazil}
\author{Rafael C. Nunes}\email{rafaelda.costa@e-campus.uab.cat}
\affiliation{Departamento de F\'{i}sica, Universitat Aut\'onoma de Barcelona, 08193, Bellaterra, Barcelona, Spain}
\affiliation{Funda\c{c}\~ao CAPES, Minist\'erio da Educa\c{c}\~ao e Cultura, 70040-020, Bras\' ilia, DF, Brazil}
\author{Everton M. C. Abreu}\email{evertonabreu@ufrrj.br}
\affiliation{Grupo de F\' isica Te\'orica e Matem\'atica F\' isica, Departamento de F\'{i}sica, Universidade Federal Rural do Rio de Janeiro, 23890-971, Serop\'edica, RJ, Brazil}
\affiliation{Departamento de F\'{i}sica, Universidade Federal de Juiz de Fora, 36036-330, Juiz de Fora, MG, Brazil}
\author{Jorge Ananias Neto}\email{jorge@fisica.ufjf.br}
\affiliation{Departamento de F\'{i}sica, Universidade Federal de Juiz de Fora, 36036-330, Juiz de Fora, MG, Brazil}

\date{\today}

\begin{abstract}

\noindent In this paper we have investigated the limits imposed by thermodynamics on a dark energy fluid. We have obtained the heat capacities and the compressibilities for a dark energy fluid.
The thermal and mechanical stabilities require these quantities to be positive. We have shown that dark energy fluids must satisfy the stability conditions and that such requirement put difficulties in the cosmic fluid models with negative constant EoS parameters. 
We have also shown that the observational constraints imposed by SN Ia, BAO and $H(z)$ data on a general dark energy fluid with a time-dependent EoS parameter are in conflict with the constraints imposed by thermodynamics. This result indicates that dark energy fluid models are unphysical. 


\end{abstract}


\maketitle

\section{Introduction}

Since the standard cold dark matter model (SCDM) was discarded by the Type Ia Supernovae observations, which point to a present day accelerated Universe \cite{acc_exp1,acc_exp2}, theoretical physicists have been challenged to find a model that agrees with the data and, at the same time, is based on a solid theoretical basis. This task is not easy. Usually, physicists start with the simplest model. Thus, the first attempt was to reintroduce the cosmological constant $\Lambda$ into Einstein's field equations. A positive $\Lambda$ term acts in the equations of motion as a constant repulsive force and therefore, can speed up the Universe at large scales. The so-called $\Lambda$CDM model is able to explain most of the current observational data and has a strong theoretical appeal since it may be linked to the zero point energy of all quantum fields filling the Universe. Physicists would have given their verdict in favor of the cosmological  constant  if it did not suffer from a serious problem: the value of the vacuum energy density obtained by observations differs from the value provided by quantum field theory by at least $60$ orders of magnitude \cite{weinberg, padmanabhan, bousso}. It is very difficult to handle this huge discrepancy. If we think it in terms of a net cosmological constant as the sum of a bare geometrical $\Lambda$ term with the quantum vacuum energy density to explain such small value, this will generate a fine-tuning problem: the absolute value of the geometrical and matter contributions to the net cosmological constant must be extremely close. Also, symmetry arguments are not enough to explain the small value of the vacuum energy observed today. The lack of a reasonable explanation for the cosmological constant problem has led physicists to explore other routes to explain the observations. The simplest one, although controversial and highly dogmatic, is to assume that vacuum energy is canceled out by some unknown symmetry in nature.  In such a scenario, scalar fields (quintessence) \cite{quintessence} stand out among the alternatives to the cosmological constant since they provide a link between particle physics and cosmology. The energy density of such a form of scalar matter must evolve with time but should mimic a cosmological constant to be compatible with the data. However, the majority of scalar fields that adjust the data have no foundations on particle physics theory and they are somewhat artificial. In fact, vacuum energy and quintessence are not the only possibilities to explain the cosmic acceleration. Fluids with negative pressure (see \cite{linder_scherrer} and references therein), k-essence \cite{k-essence}, phantom fields \cite{phantom}, modifications of gravity theory \cite{nojiri}, brane worlds \cite{brane} and other open-minded propositions that do not require additional sources for the energy contend of the universe \cite{opened-minded} are also in the game. %

Collectively, all models that invokes additional sources of energy to explain the cosmic acceleration are called dark energy (DE) (see \cite{copeland} for a review).
The easiest way to implement dark energy is through a dark energy fluid. By assuming that general relativity is the correct theory of gravitation,  the pressure of a DE fluid must be sufficiently negative to make the sum $\sum_i(\rho_i +3p_i/c^2)$ negative in order to produce an accelerated expansion of the Universe. Dark energy fluids are frequently characterized by the equation of state (EoS) parameter (the ratio between its pressure and its energy density) $w=p_{{\rm DE}}/\rho_{{\rm DE}} c^2$ which can be constant or time-dependent. Such a phenomenological approach mimics the vacuum energy\footnote{It should be stressed that the link between the zero point energy of all matter fields filling the Universe and the cosmological constant comes from a semi-classical quantization process where quantum effects are taken into account only in the energy-momentum tensor. Since a fluid with EoS $p=-\rho c^2$ acts in the field equations in the same way as the zero point energy contribution, vacuum energy is commonly termed as dark energy. But, in spite of its mathematical equivalence, a dark energy fluid with a EoS $p=-\rho c^2$ is physically distinct from vacuum energy.} ($w=-1$), scalar fields ($-1\le w(t)\le 1$), phantom fields ($w(t)<-1$) and many other forms of exotic matter. Although quintessential ($-1\le w(t)\le 1$) and phantom fluids ($w(t)<-1$) act, respectively, in the same way as the canonical and phantom scalar fields in Einstein's equations, it should be stressed that fluids and fields are two  physically different things. For instance, while the sound speed of a dark energy fluid $c_s^2=\partial p_{{\rm DE}}/\partial\rho_{{\rm DE}}$ can evolve with time, the sound speed of a quintessence (phantom) scalar field is always equal to $c^2$ ($-c^2$) \cite{linder_scherrer, mukhanov}. In fact, if the acceleration of the Universe is due to some type of dark energy, the great issue that must be answered is: what is dark energy? It is the vacuum energy, some type of scalar matter or even some form of more exotic matter?  Observational data are not enough to allow us to decide between different kinds of dark energy since most of the proposed models are able to adjust the data seamlessly. We therefore need to go deeper into theory in order to achieve a better understanding of the mechanisms behind the cosmic acceleration. 
In this direction, thermal physics is of particular importance in dark energy studies. The laws of thermodynamics are based on experimental evidence and they apply to all types of macroscopic systems. Unlike classical mechanics or electromagnetism, thermodynamics does not predict specific numerical values for observables. Thermodynamics sets limits on physical processes. The power of thermodynamics resides in its generality. Therefore, to explore the thermodynamical behavior of the cosmic fluids that pervade our Universe may be a line of attack for unveiling the nature of the content of the Universe, particularly the hypothetical DE fluid. For example, the positiveness of the entropy may be one of the main weapons to impose bounds on the EoS parameter of dark energy fluids\cite{entropy_positiveness}.   To apply the laws of thermodynamics to the dark energy fluid theory can help us to constrain, or even to rule out some dark energy fluid models. Below we will carry out this task.

\section{Thermodynamics of the cosmic fluids}

Let us consider an expanding, homogeneous and isotropic Universe filled by (baryonic and dark) matter, described by a pressureless perfect fluid ($w=0$), radiation, described by a perfect fluid with a EoS parameter $w=1/3$ and dark energy, described by a perfect fluid with a EoS parameter $w=p/\rho c^2$. Homogeneity and isotropy imply that all physical distance scales with the same factor $a(t)$, called the scale factor of the Universe. Thus, the physical volume of the Universe at a given time is $V=a^3(t)V_0$\footnote{Here the index $0$ will denote the present time value of an observable and we will adopt the convention $a_0=1$}. In such a model the internal energy of the $i$-th fluid  component can be written as
\begin{equation}
\label{FRW_energy}
U_i=\rho_i c^2 V.
\end{equation}

\noindent Assuming a reversible adiabatic expansion, 
the first law of thermodynamics
\begin{eqnarray}
\label{Gibbs_law}
T_idS_i&=&dU_i+p_idV,
\end{eqnarray}

\noindent leads to the so-called fluid equation
\begin{equation}
\label{fluid_eq}
d\ln\rho_i+3(1+w_i)d\ln a=0,
\end{equation}

\noindent which expresses the energy-momentum conservation. Assuming that density is a function of both temperature and volume, i. e., $\rho_i=\rho_i(T_i, V)$, the fact that $dS_i$ is an exact differential implies that \cite{weinberg1971}
\begin{equation}
\label{temperature_law_w}
d\ln T_i=-3w_id\ln a,
\end{equation}

\noindent or, using (\ref{fluid_eq}) to eliminate $w_i$,
\begin{equation}
\label{temperature_law_rho}
d\ln T_i=d\ln \rho_i+ 3d\ln a.
\end{equation}

\noindent Integrating the temperature law (\ref{temperature_law_rho}) we obtain that
\begin{equation}
\label{T_rho_rel}
\frac{T_i}{T_{i,0}}=\frac{\rho_i}{\rho_{i,0}}a^3
\end{equation}

\noindent or, in a more suggestive form 
\begin{equation}
\frac{1}{w_i}\frac{p_i V}{T_i}=\frac{1}{w_{i,0}}\frac{p_{i,0} V_0}{T_{i,0}}={\rm constant}.
\end{equation}

\noindent The above equation generalizes the ideal gas law for a time dependent EoS parameter. Finally, the fluid energy can be written in terms of the temperature as
\begin{equation}
\label{internal_energy}
U_i=U_{i,0}\frac{T_i}{T_{i,0}}.
\end{equation}






%
In what follows, we will derive the expressions for the heat capacity, compressibility and the thermal expansibility. These thermodynamical derivatives are easily accessible experimentally concerning any terrestrial fluid. The heat capacity and the compressibility of the fluid are related, respectively, to both  thermal stability and mechanical stability and it must be greater than zero since the stable equilibrium has been reached by the system. Thus, we can use these variables to impose bounds on the EoS parameter of DE.

\subsection{The Universe's heat capacity}

The classical thermodynamical definition of a fluid's heat capacity $C_i$ is \cite{callen}
\begin{equation}
\label{SH_definition}
dQ_i=C_idT_i,
\end{equation} 

\noindent where $dT_i$ is the fluid temperature increase due to an absorbed heat $dQ_i=T_idS_i$. The heat capacity of a fluid will differ depending on whether the fluid is heated at constant volume or at constant pressure. From the first law of thermodynamics, Eq.  (\ref{Gibbs_law}), at constant volume, (\ref{SH_definition}) becomes
\begin{equation}
\label{CV0}
dU_i=C_{iV}dT_i,
\end{equation}

\noindent where
\begin{equation}
\label{CV}
C_{iV}=\Big(\frac{\partial U_i}{\partial T_i}\Big)_V,
\end{equation}

\noindent is the fluid's heat capacity at constant volume. The heat capacity at constant pressure can be built up from the enthalpy
\begin{equation}
\label{enthalpy}
h_i=U_i+p_iV,
\end{equation}

\noindent in terms of which the first law of thermodynamics is written as
\begin{equation}
\label{1st_law_enthalpy}
dQ_i=dh_i-Vdp_i.
\end{equation}

\noindent Thus, at constant pressure, (\ref{SH_definition}) becomes
\begin{equation}
\label{CP0}
dh_i=C_{p_i}dT_i,
\end{equation}

\noindent where
\begin{equation}
\label{CP}
C_{p_i}=\Big(\frac{\partial h_i}{\partial T_i}\Big)_{p_i},
\end{equation}

\noindent is the fluid's heat capacity at constant pressure.

From equation (\ref{internal_energy}), it is easy to show that
\begin{equation}
\label{CV_cosmic}
C_{iV}=\frac{U_{i,0}}{T_{i,0}}={\rm constant},
\end{equation}

\noindent for any component of the Universe. Since $p_iV=w_iU_i$, the enthalpy (\ref{enthalpy}) becomes 
\begin{equation}
h_i=(1+w_i)U_i
\end{equation}
\noindent and, from equations (\ref{internal_energy}) and (\ref{temperature_law_w}), we have 
\begin{equation}
\label{Cp_time_dependent}
C_{p_i}=\Big(1+w_i-\frac{1}{3}\frac{d\ln \vert w_i\vert}{d\ln a}\Big)C_{iV}.
\end{equation}

\noindent Since $U_{i,0}=\rho_{i,0}c^2V_0$, the specific heat (the heat capacity per mass unit) at constant volume is
\begin{equation}
\label{specific_heat}
c_{iV}\equiv\frac{C_{iV}}{\rho_{i,0}V_0}=\frac{c^2}{T_{i,0}}.
\end{equation}

\noindent For relativistic matter $T_{r,0}=2.725\,{\rm K}$ so that $c_{rV}$ and $c_{p_r}$ are of order of $\sim10^{13}\,{\rm cal}\cdot{\rm g}^{-1}\cdot{\rm K}^{-1}$. Since the temperature of the other components must be smaller than the temperature of relativistic matter, the specific heat of the relativistic matter is an inferior limit for the Universe's specific heat. As expected, this result reveals that the Universe is a huge thermal reservoir. Unfortunately, despite easy experimental access for terrestrial fluids, we cannot isolate a cosmologically significant portion of the Universe, to provide an enormous amount of heat, and to measure the temperature change of our Universe sample to obtain its specific heat experimentally. 

\subsection{Compressibility and expansibility}

\noindent If we consider the volume as function of temperatures and pressures we have that\footnote{Remember that we are assuming that the fluids evolve separately, that is, they do not exchange heat, as shown by eq. (\ref{fluid_eq})}
\begin{equation}
\label{vol_diff}
dV=\sum_i\Big[\Big(\frac{\partial V}{\partial T_i}\Big)_{p_i}dT_i+\Big(\frac{\partial V}{\partial p_i}\Big)_{T_i}dp_i\Big].
\end{equation}

\noindent We will now define the thermal expansivity, which measures the volume thermal expansion at constant pressure, by
\begin{equation}
\label{alpha}
\alpha_i=\frac{1}{V}\Big(\frac{\partial V}{\partial T_i}\Big)_{p_i},
\end{equation}

\noindent and the isothermal compressibility,  which measures the relative change of volume with increasing pressure at fixed temperature, by 
\begin{equation}
\label{kappa_t}
\kappa_{T_i}=-\frac{1}{V}\Big(\frac{\partial V}{\partial p_i}\Big)_{T_i}.
\end{equation}

\noindent Analogously to the isothermal compressibility, we can define the adiabatic compressibility $\kappa_{S_i}$, if, instead of temperature, the entropy is kept fixed. 
%
%
\noindent It can be shown that the isothermal compressibility and the isothermal expansibility are connected by
\begin{equation}
\label{ak_ratio}
\frac{\alpha_i}{\kappa_{T_i}}=\Big(\frac{\partial p_i}{\partial T_i}\Big)_V,
\end{equation}

\noindent and that the ratio between the adiabatic and the isothermal compressibilities are equal to the ratio between the  heat capacities at constant volume and at constant pressure, i.e.,
\begin{equation}
\label{kk_ratio}
\frac{\kappa_{S_i}}{\kappa_{T_i}}=\frac{C_{iV}}{C_{p_i}}.
\end{equation}

\noindent Noting that $p_iV=w_iC_{iV}T_i$ and using (\ref{temperature_law_w}) we obtain
\begin{equation}
\label{a_time_dependent}
\alpha_i=\frac{C_{iV}}{p_iV}\Big(w_i-\frac{1}{3}\frac{d\ln \vert w_i\vert}{d\ln a}\Big).
\end{equation}

\noindent From (\ref{ak_ratio}) it is easy to show that
\begin{equation}
\kappa_{T_i}=\frac{\alpha_i V}{w_iC_{iV}},
\end{equation}

\noindent and from the above equation and (\ref{kk_ratio}) we have 
\begin{equation}
\kappa_{S_i}=\frac{\alpha_i V}{w_iC_{p_i}}.
\end{equation}


\subsection{Stability conditions}

A thermodynamic system involving only the work due to the hange in volume will be in stable equilibrium if the second order variation 

\begin{equation}
\label{stable_eq}
\delta^2 U_i=\delta T_i\delta S_i - \delta p_i\delta V
\end{equation}

\noindent is greater than zero \cite{kubo}. Otherwise, a thermodynamic stability is not obtained. If $T_i$ and $V$ are the independent variables, it is easy to show that

\begin{equation}
\label{stable_eq_TV}
\delta^2 U_i=\frac{C_{iV}}{T_i}\delta T_i^2+\frac{1}{V\kappa_{T_i}}\delta V^2. 
\end{equation}
  
\noindent Also, taking $S_i$ and $p_i$ as independent variables, Eq. (\ref{stable_eq}) becomes

\begin{equation}
\label{stable_eq_Tp}
\delta^2 U_i= \frac{T_i}{C_{p_i}}\delta S_i^2+V\kappa_{S_i}\delta p_i^2\,\,.
\end{equation}

\noindent It is easy to see that, if a given cosmic fluid component reach the thermodynamic stability $\delta^2U_i\ge0$, Eqs. (\ref{stable_eq_TV}) and (\ref{stable_eq_Tp}) imply that

\begin{equation}
\label{stability_conditions}
C_{iV},\,C_{p_i},\,\kappa_{T_i},\, \kappa_{S_i}\ge 0
\end{equation}

\noindent simultaneously. Conversely, if stability is not reached, i. e., $\delta^2U_i<0$, the heat capacities and the compressibilities are all negative simultaneously. According to (\ref{CV_cosmic}), $C_{iV}$ is constant and positive for any fluid component, showing that the Universe components, viewed as non interacting perfect fluids, are necessarily constrained by (\ref{stability_conditions})\footnote{It should be stressed that inhomogeneous systems or interacting systems can have a negative heat capacity. Examples of inhomogeneous systems are Globular clusters \cite{globular} and black holes \cite{BH} whose density is high in the center due gravitational field strength.}. Positiveness of the heat capacity are related to thermal stability and positiveness of compressibility are related to mechanical stability. Additionally, it can be shown that $C_{iV},C_{p_i},\kappa_{S_i},\,{\rm and}\,\kappa_{T_i}$ are related by 
\begin{equation}
C_{p_i}=C_{iV}+\frac{TV\alpha_i^2}{\kappa_{T_i}}
\end{equation}
\noindent and
\begin{equation}
\kappa_{T_i}=\kappa_{S_i}+\frac{TV\alpha_i^2}{C_{p_i}}.
\end{equation}
\noindent Thus, since the system satisfies the stability conditions (\ref{stability_conditions}), we have that
\begin{equation}
\label{stability_conditions_2}
C_{p_i}\ge C_{iV}\, \mbox{and}\, \kappa_{T_i}\ge\kappa_{S_i}. 
\end{equation}

  
\section{Constraints on dark fluids}

From (\ref{Cp_time_dependent}) and (\ref{a_time_dependent}) it is easy to see that the conditions (\ref{stability_conditions}) and (\ref{stability_conditions_2}) are satisfied only if the fluid EoS parameter obeys the constraint
\begin{equation}
\label{thermo_bound}
w_i-\frac{1}{3}\frac{d\ln \vert w_i\vert}{d\ln a}\ge0.
\end{equation}

\begin{table}[h!]
\begin{center}
\begin{tabular}{ccc}
\hline
\hline
\\ 
Reference & $w$ & $w-\frac{1}{3}\frac{d\ln \vert w\vert}{d\ln a}\ge0\quad\forall a\in[0,\infty)$ \\ 
\\
\hline
\hline   
\\
\cite{Wet}&$\frac{w_0}{(1-b\ln a)^2}$& No\\ 
\\
\cite{LH}&$w_f+\frac{\Delta w a_t^{1/\tau}}{a_t^{1/\tau}+a^{1/\tau}}$& No\\
\\
\cite{HM}&$w_fw_i\frac{a^l+a_t^l}{w_ia^l+w_fa_t^l}$& No\\
\\
\cite{JBP} & $w_0+w'_0(a-a^2)$            & No \\
\\                          
\cite{BA1} & $w_0+w'_0\frac{a-1}{1-2a+2a^2}$  & No \\
\\
\cite{BA2} & $w_0+w'_0\frac{a^{\beta}-1}{\beta}$        &  No\\
\\
\hline
\hline

\end{tabular}
\caption{Thermodynamic viability of some DE parametric models found in literature.Here, $w'_0=(dw/da)_{a=1}$.\label{Tab:1}} 
\end{center}
\end{table}

\noindent It is obvious that, if $w_i$ is constant, thermal and mechanical stability implies that $w_i\ge0$. This result implies that homogeneous negative pressure fluids with a constant EoS parameter are unphysical. This does not mean that vacuum energy cannot be the piece behind cosmic acceleration. As we have already stressed, although mathematically equivalent, physically a fluid with pressure $p=-\rho c^2$ is quite different from vacuum energy. Therefore, if the accelerated expansion of the Universe is caused by a dark energy fluid, its EoS parameter must be time-dependent. In Table \ref{Tab:1} we list some time-dependent guesses of the DE EoS parameter. Although in excellent agreement with the data, these phenomenological models do not satisfy the thermodynamical bound (\ref{thermo_bound}). 

Now, let $w$ and $\rho_{{\rm DE}}$ denotes, respectively, the EoS parameter and the density of the dark energy fluid. Since $w$ must evolves with the time, we can use (\ref{fluid_eq}) to rewrite the inequality (\ref{thermo_bound}) as 
\begin{equation}
 3+\frac{d\ln \rho_{{\rm DE}}}{d\ln a}\le-\frac{d\ln \vert w\vert}{d\ln a}.
\end{equation}

\noindent Integrating both sides of the inequality above we can write that
\begin{equation}
\label{EoS_temp_constraint}
\vert\frac{w}{w_{0}}\vert\le\frac{\rho_{{\rm DE},0}}{\rho_{{\rm DE}}a^3}=\frac{T_{{\rm DE},0}}{T_{{\rm DE}}},
\end{equation}

\noindent where we have used (\ref{T_rho_rel}). According to (\ref{temperature_law_w}) the temperature of the fluid will increase (decrease) with the expansion of the Universe if $w<0\,(w>0)$. The constraint (\ref{EoS_temp_constraint}) reveals that an eternal accelerated expansion cannot be sustained by a DE fluid. If $w<0\quad\forall\, a\ge1$, $T_{{\rm DE}}(a\to\infty)\to\infty$ and $\vert w(a\to\infty)\vert\to 0$, which means that the accelerated expansion will stop in the distant future. On the other hand, if the sign of $w$ changes in the course of the expansion, the Universe can enter and leave in an accelerated expansion phase which depends on how many times the sign of $w$ changes but it cannot keep an acceleration expansion phase forever. This transient behavior imposed by thermodynamics is particularly important for the formulation of the String/M theory since an eternal accelerated expansion implies that a conventional S-matrix cannot be built \cite{Fischler, Hellerman,Halyo}.

Now, at the present time the inequality (\ref{thermo_bound}) becomes
\begin{equation}
\label{present_day_TB}
w'_0\ge 3w_0^2.
\end{equation}

\noindent In order to check the compatibility of the observational data with the above thermodynamic constraint we follow the approach developed in \cite{BA3} which is one of the less model dependent methods to probe the DE EoS time-dependence. This approach consists of assuming that the DE density admits a Taylor expansion in the range $(\tilde{a}-\epsilon_-,\tilde{a}+\epsilon_+)$, that is, 
\begin{eqnarray}
\rho_{{\rm DE}}(a)&=&\rho_{{\rm DE}}(\tilde{a})+\frac{d\rho_{{\rm DE}}}{da}\Big|_{a=\tilde{a}}(a-\tilde{a})+\nonumber\\
&+&\frac{1}{2}\frac{d^2\rho_{{\rm DE}}}{da^2}\Big|_{a=\tilde{a}}(a-\tilde{a})^2+\cdots
\end{eqnarray}

\noindent and then use the conservation equation (\ref{fluid_eq}) as a recurrence formula to write the derivatives of $\rho_{{\rm DE}}$ in terms of the derivatives of $w$, i. e.,
\begin{eqnarray}
\frac{d\rho_{{\rm DE}}}{da}&=&-\frac{3}{a}(1+w)\rho_{DE},\nonumber\\
\frac{d^2\rho_{{\rm DE}}}{da^2}&=&\Big[\frac{3}{a^2}(1+w)+\frac{9}{a^2}(1+w)^2-\frac{3}{a}\frac{dw}{da}\Big]\rho_{DE},\nonumber\\
\vdots\quad&&\vdots .\nonumber
\end{eqnarray}

\noindent This approach allows us to constrain $w$ and its derivatives at different redshifts simply by changing the series expansion center $\tilde{a}$. Since for sufficiently small values of $\epsilon_{\pm}$ the second order approximation must work reasonably well, we will restrict our analysis up to the second order expansion of $\rho_{{\rm DE}}$ around $\tilde{a}=a_0=1$\footnote{For example, taking the expansion center $\tilde{a}=a_0=1$ and $\epsilon_+=1/3$ it is possible to cover the redshift range $0\le z\le 2$.}. Thus, by choosing the expansion center at $a_0$, the second order approximation of the DE density becomes

\begin{eqnarray}
\label{DE_density_exp}
\rho_{{\rm DE}}(a)&=&\rho_{{\rm DE},0}\Big\{1+3(1+w_0)(1-a)+\frac{1}{2}\big[3(1+w_0)\nonumber\\&-&3w'_0+9(1+w_0)^2\big](1-a)^2\Big\}.
\end{eqnarray}

\begin{figure}[t]
\includegraphics[width=3.0truein,height=3.0truein]{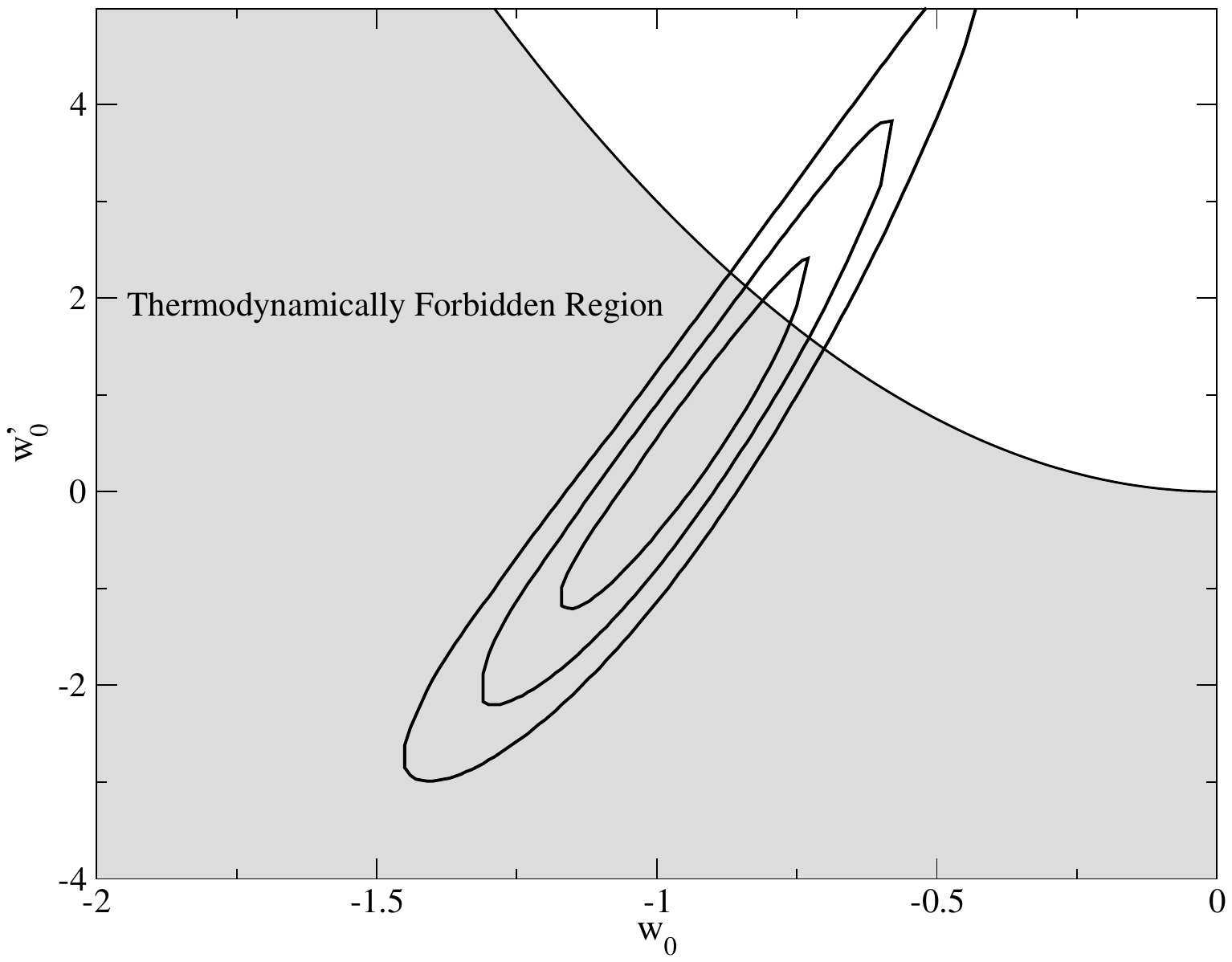}
\caption{The $w_0-w'_0$ parametric space. The thermodynamically forbidden region corresponds to points in the space of phase for which the inequality (\ref{present_day_TB}) is not satisfied. The contours are drawn for $\Delta \chi^2 = 2.30$, 6.17 and 11..\label{fig:1}}
\end{figure}

\noindent Figure \ref{fig:1} shows the observational constraints in 1, 2 and 3 $\sigma$ on $w_0$ and $w'_0$ for a specially flat, homogeneous and isotropic Universe filled by relativistic matter, non-relativistic matter and  a DE  fluid described by (\ref{DE_density_exp}) obtained from 580 supernovae data of Union 2.1 compilation, the six estimates of the BAO points given in Table 3 of Ref.~\cite{blakea} and the $28$ measurements of the Hubble function $H(z)$ compiled by Liao {\it et al.} \cite {hdata} and  Farooq \& Ratra \cite{hdata1} (see also the references therein). The present value of the Hubble parameter $H_0$ and the matter density parameter $\Omega_{m,0}$ were marginalized. For this data combination,  the best fit values are $w_0=-0.96^{+0.22}_{-0.21}$ and $w'_0=-0.33^{+2.00}_{-1.53}$, with the upper and lower values denoting the one parameter $1\sigma$ errors. 
As we can see, a large portion of the $w_0-w'_0$ confidence regions lies in the unphysical region. This lack of sensitivity of the data to the physical constraint $w'_0\ge3w_0^2$ can be interpreted as an evidence against the DE fluid models since if a DE fluid is causing the accelerated expansion the data would not be in conflict with its physical properties. However, if DE cannot be described as a fluid with negative pressure the data should not be expected to follow the physical properties of such a fluid, but would rather force the fluid's parameters to converge to the values that better approximate the true mechanism behind the cosmic acceleration regardless of the  physical bounds that a hypothetical DE fluid must obey. 

\section{Final Remarks}

What is causing the accelerated expansion of the Universe? Is it DE or it is because Einstein's general relativity does not work at large scales? Physicists  have worked on both fronts to answer that question.  
Particularly, on the DE side, a large number of models has been proposed. 
In this article we believe that we have taken a big step in the understanding of the current phase of accelerated expansion of the Universe. We have studied the thermodynamical aspects of an expanding, homogeneous and isotropic Universe filled by matter (baryonic plus dark), relativistic matter (radiation plus neutrinos) and a hypothetical DE. By regarding the cosmic components as perfect fluids, we have estimated the Universe specific heat and examined the constraints imposed by classical thermodynamics on the DE fluid. We have shown that the cosmic fluids necessarily must  reach the thermodynamic stability. Such a requirement implies that negative pressure perfect fluids with a constant EoS parameter are unphysical. We have also shown that the observational constraints on a DE fluid with a time-dependent EoS parameter are in conflict with the physical constraints imposed by thermodynamics. This result suggests that to add a DE fluid to the content of the Universe may not be the answer to the cosmic acceleration problem.

Although our analysis implies that a DE fluid with a EoS parameter $w=-1$ is unphysical, vacuum energy remains untouched since it is physically different from a DE fluid with $p=-\rho c^2$.  If we ignore the cosmological constant problem by setting $\rho_{\Lambda}=0$, then the cosmic acceleration due to some type of dark energy fluid does not seem to be a good way to address the problem. By doing so, we are replacing a physically well motivated explanation and considering a hypothesis that, at least in its simplest formulation, cannot be corroborated by the basic physical laws. 
However, a question still remains: is the cosmological constant the explanation for the accelerated expansion? The $\Lambda$-term certainly is the simplest solution but no one can guarantee that it is the true answer. Scalar fields and other  models that do not require any additional sources remain as possibilities. Thus, to find out deviations of the cosmological term will remain as one of the hottest theoretical investigation lines concerning cosmic acceleration. If DE fluids are out of the game, approaches such as the kinematic method developed in \cite{edesio} can be a useful tool to search for such deviations. 


\begin{acknowledgments}

R.C.N. acknowledges financial support from CAPES Scholarship Box 13222/13-9.  E.M.C.A. thanks CNPq (Conselho Nacional de Desenvolvimento Cient\'{i}�fico e Tecnol\'ogico), a brazilian scientific support federal agency, for  partial financial support, grants numbers 301030/2012-0 and 442360/2014-0. The authors are very grateful to Thomas Dumelow for a critical reading of the manuscript.

\end{acknowledgments}

\end{document}